%% file: AEL.tex
\documentclass[12pt, letterpaper]{article}
\usepackage[margin=1in]{geometry}
\usepackage{amsmath, amssymb, booktabs, graphicx, natbib, parskip, setspace, tikz, subcaption}
\makeatletter
\newcommand\footnoteref[1]{\protected@xdef\@thefnmark{\ref{#1}}\@footnotemark}
\makeatother

\title{
    The Beginning of the Trend: \\
    Interest Rates, Profits, and Markups
}

\author{
    Anton Bobrov \\
    James Traina%
    \thanks{
        Bobrov: Federal Reserve Bank of San Francisco, anton.bobrov@sf.frb.org.
        Traina: Federal Reserve Bank of San Francisco, traina@uchicago.edu.
        The views here are those of the authors and not those of the Federal Reserve Bank of San Francisco or the Federal Reserve System. We thank Brian Albrecht, John Fernald, Marianna Kudlyak, Sanjay Singh, and Uyen Tran for insightful discussions.
    }
}

\date{October 2023}

\begin{document}

\maketitle

\begin{abstract}
    \noindent
    Recent highly cited research uses time-series evidence to argue the decline in interest rates led to a large rise in economic profits and markups. We show the size of these estimates is sensitive to the sample start date: The rise in markups from 1984 to 2019 is 14\% larger than from 1980 to 2019, a difference amounting to a \$3000 change in income per worker in 2019. The sensitivity comes from a peak in interest rates in 1984, during a period of heightened volatility. Our results imply researchers should justify their time-series selection and incorporate sensitivity checks in their analysis.
\end{abstract}

{\medskip \footnotesize
}

\clearpage

Since the early 1980s, the labor share of value-added and the real interest rate have declined significantly. These declines have led to a debate about the magnitude of an implied rise in market power \citep{barkai2020declining, farhi2018accounting, karabarbounis2019accounting, davis2023profit}. As a leading example, \citet{barkai2020declining} delineates these trends from 1984 to 2014: The decline in interest rates led to a 5 pp decline in the cost of capital and a 7 pp decline in the capital share; the implied economic profit share increased by 14 pp, amounting to over \$1 trillion in economic profits in 2014. 

However, there is substantial disagreement about the growth of economics profits and markups, with different methodologies and samples leading to disparate magnitudes. \citet{basu2019} finds markups measured using aggregate data in 1984-2014 \citep{barkai2020declining}, industry data in 1988-2015 \citep{hall2018new}, and firm data in 1980-2016 \citep{de2020rise} produce a wide range of estimates. In this paper, we ask: How does sample selection influence the estimated growth in market power?

Our paper shows time-series selection heavily influences the economic size of these results, challenging their robustness.\footnote{Time-series selection sensitivity is one of many other time-series inference traps in research. Select examples include: \citet{card1995time} on publication bias; \citet{white2000reality} on multiple comparisons, and the large literature on the factor zoo, as recently discussed in \citet{jensen2022there}.} We contribute to the market power literature by quantifying the role of sample selection on the estimated trends implied by decreasing interest rates. Extending our 4-decade sample back by just 4 years explains 25\% of the trend in the real interest rate, 37\% in the cost of capital, 14\% in the profit share, and 14\% in the aggregate markup. This sensitivity amounts to a difference of \$260 billion of economic profits in 2019.

More generally, we show sensitivity is pronounced when analyzing any trend using high-volatility time series, as with real interest rates in the 1980s. Data points far from the time value mean or the dependent variable mean have higher influence on time-series regression results. Hence, points at the beginning or end of a sample are mechanically influential, and points drawn from a highly volatile time series are more likely influential. Therefore, the volatility in financial markets poses a challenge in accurately quantifying trends. Our empirical contribution indicates this challenge is plausibly significant, calling for justified sample selection and robustness checks.

We emphasize the application to quantifying secular trends in market power for two reasons. First, researchers use these different magnitudes to estimate models and inform policies. Following Executive Order 14036, the \citet{cea2023} uses evidence from \citet{shapiro2018antitrust} and \citet{philippon2019causes-key} to reorient competition policy. Precise measurements of economic profits and markups are also key targets for modeling and welfare \citep{edmond2023costly, farhi2018accounting, eggertsson2021kaldor}. For instance, \citet{blanchard2019public} cites the scale of rising profits to argue the cost of public debt is small. Second, as highlighted, there is a substantial disagreement about the growth of markups.\footnote{Our analysis therefore contributes to ongoing empirical research on secular trends in the US economy. Examples include the rise of national-level concentration \citep{covarrubias2020good}, superstar firms \citep{autor2020fall}, and measured public-firm markups \citep{de2020rise, traina2018aggregate}.} We clarify one major reason for these disparate magnitudes: Sample selection.

\section{Interest Rates, Profits, and Markups}

We employ the standard methodology of \citet{hall1967tax} to measure the cost of capital using financial-market rates, which track the decline in 10-year Treasury yields \citep{barkai2020declining, karabarbounis2019accounting, davis2023profit}. We then measure profits and markups as the residual from national income less payments to labor and capital. Our measurement follows \citet{barkai2020declining}: The main variables come from the Bureau of Economic Analysis’s National Income and Product Accounts Table 1.14 (value added, labor compensation), Fixed Assets Accounts Table 4.1 (capital, depreciation, and inflation), and Integrated Macroeconomic Accounts Table S.5.a (inventories); tax rates come from the Organisation for Economic Cooperation and Development and the Tax Foundation.

Our approach calculates the change in the estimated linear trend through 2019 for different start dates in the 1980s. For each time series, we estimate the simple linear regression:
    $$y_t = \beta_t + \alpha + \varepsilon_t.$$
In the left panels, we plot the entire time series at the highest frequency available on the Federal Reserve Economic Data database. In the right panels, we plot the percentage change in the trend for every starting year in the 1980s compared to a base year of 1980.

\begin{figure}[!htbp]
    \centering
    \begin{subfigure}[t]{0.49\textwidth}
        \includegraphics[width=3.25in]{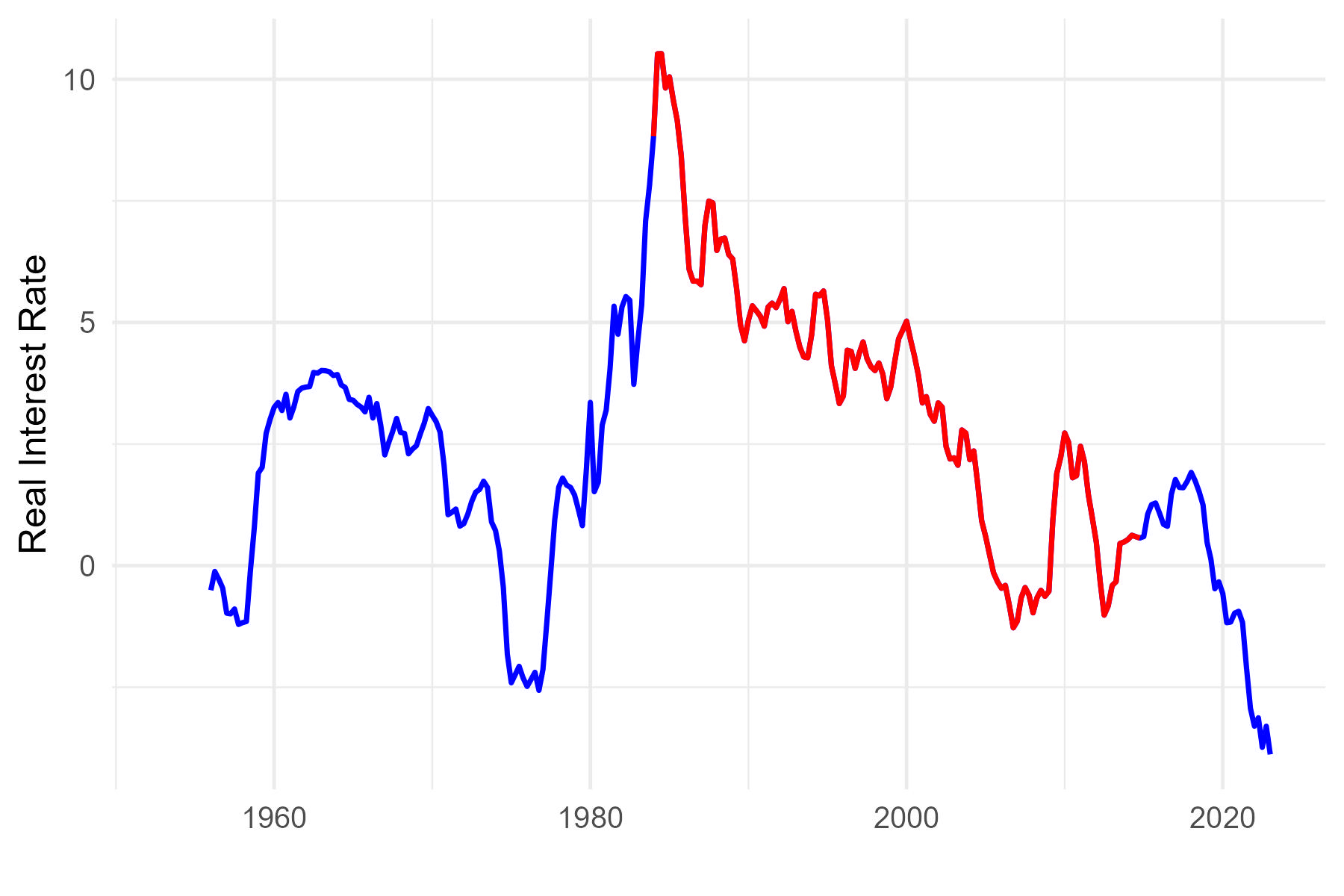}
        \caption{Real Interest Rate}
        \label{fig:fig1a}
    \end{subfigure}
    \begin{subfigure}[t]{0.49\textwidth}
        \includegraphics[width=3.25in]{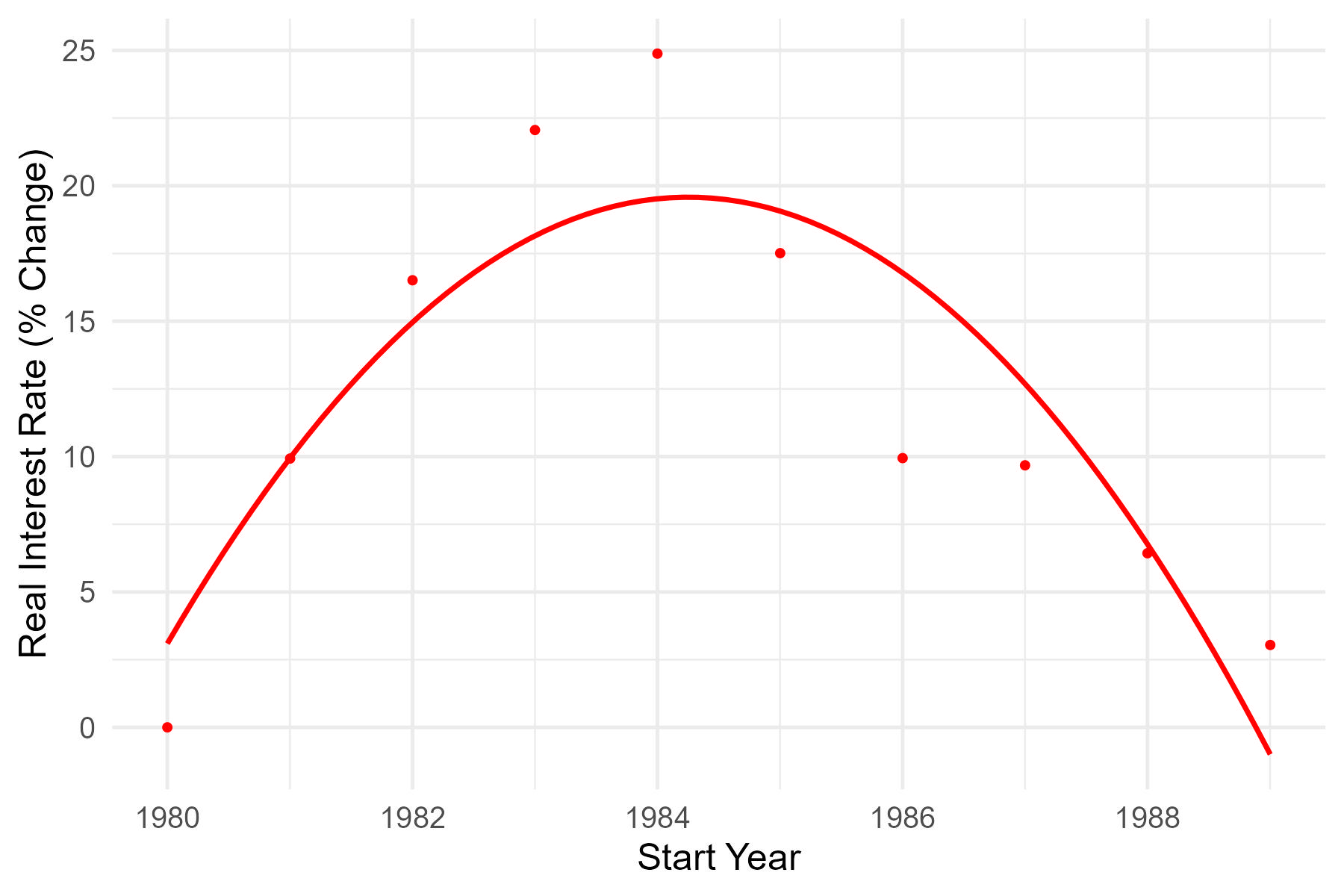}
        \caption{Estimated Trend}
        \label{fig:fig1b}
    \end{subfigure}
    \caption{Real Interest Rate: $i - \nu$. The real interest rate is measured as the market yield on U.S.10-Year Treasury Securities less expected capital inflation, constructed as the three-year moving average of realized capital inflation from the BEA. Shown at quarterly frequency. }
    \label{fig:fig1}
\end{figure}

Figure \ref{fig:fig1a} displays the real interest rate. The portion in red is from 1984 to 2014, the period commonly referenced in the literature. Figure \ref{fig:fig1b} plots the percentage change in the linear trend of the real interest rate, starting from different years. We fit a quadratic function over these points.

The linear trend for the real interest rate is steepest when starting the series in 1984, which is the most commonly cited start date in the literature. Data points from the beginning of the sample are highly influential. The real interest rate was 7 pp lower in 1980 than in 1984. The peak in the real interest rate occurred in 1984.

The \citet{hall1967tax} cost of capital $R_c$ is:
    $$R_c = \left( \rho - \nu + \delta \right)\frac{1 - z \tau}{1- \tau}.$$
The \citet{hall1967tax} formula maps the cost of finance\footnote{The cost of finance is also known as the weighted average cost of capital (WACC) in financial economics.} to the cost of capital used in real investment decisions by adjusting it for expected capital inflation $\nu$, depreciation $\delta$, taxes $\tau$, and depreciation allowances $z$. The formula is widely used in the capital investment literature because it holds in a large class of models, relying only on a simple arbitrage argument. Here, $\rho$ is the cost of finance and is defined as $dR_d +(1-d)R_d$, where $d$ is the debt share of financial assets, $R_d$ is the expected return on debt measured as the Moody's Baa bond yield, and $R_e$ is the expected return on equity measured as the 10-year Treasury yield plus a 5\% risk premium. \citet{barkai2020declining}, \citet{karabarbounis2019accounting}, and \citet{davis2023profit} all show the cost of finance $\rho$ closely tracks the interest rate $i$; we focus on $\rho$ to make it easy to compare with \citet{barkai2020declining}.

\begin{figure}[!htbp]
    \centering
    \begin{subfigure}[t]{0.49\textwidth}
        \includegraphics[width=3.25in]{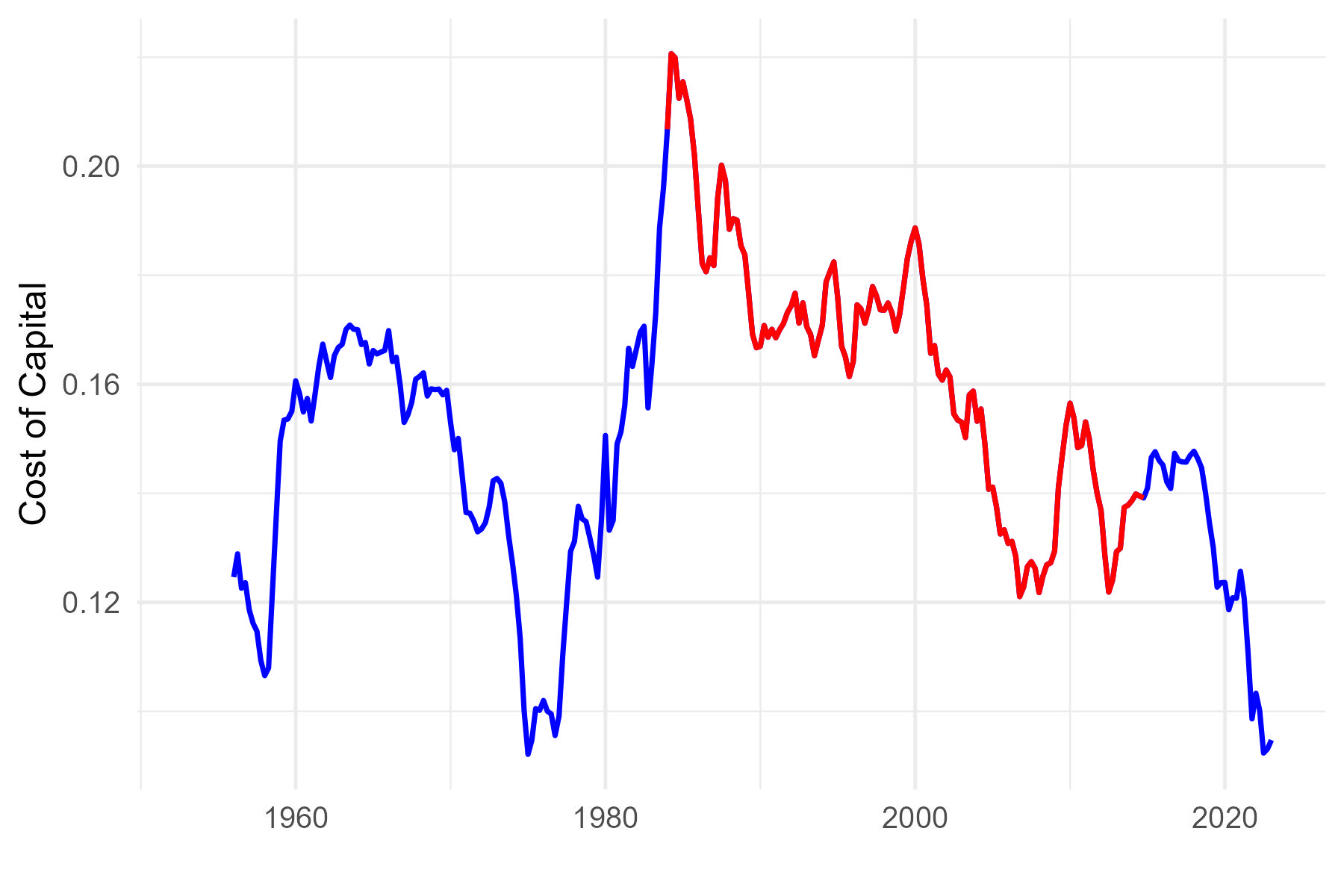}
        \caption{Cost of Capital}
        \label{fig:fig2a}
    \end{subfigure}
    \begin{subfigure}[t]{0.49\textwidth}
        \includegraphics[width=3.25in]{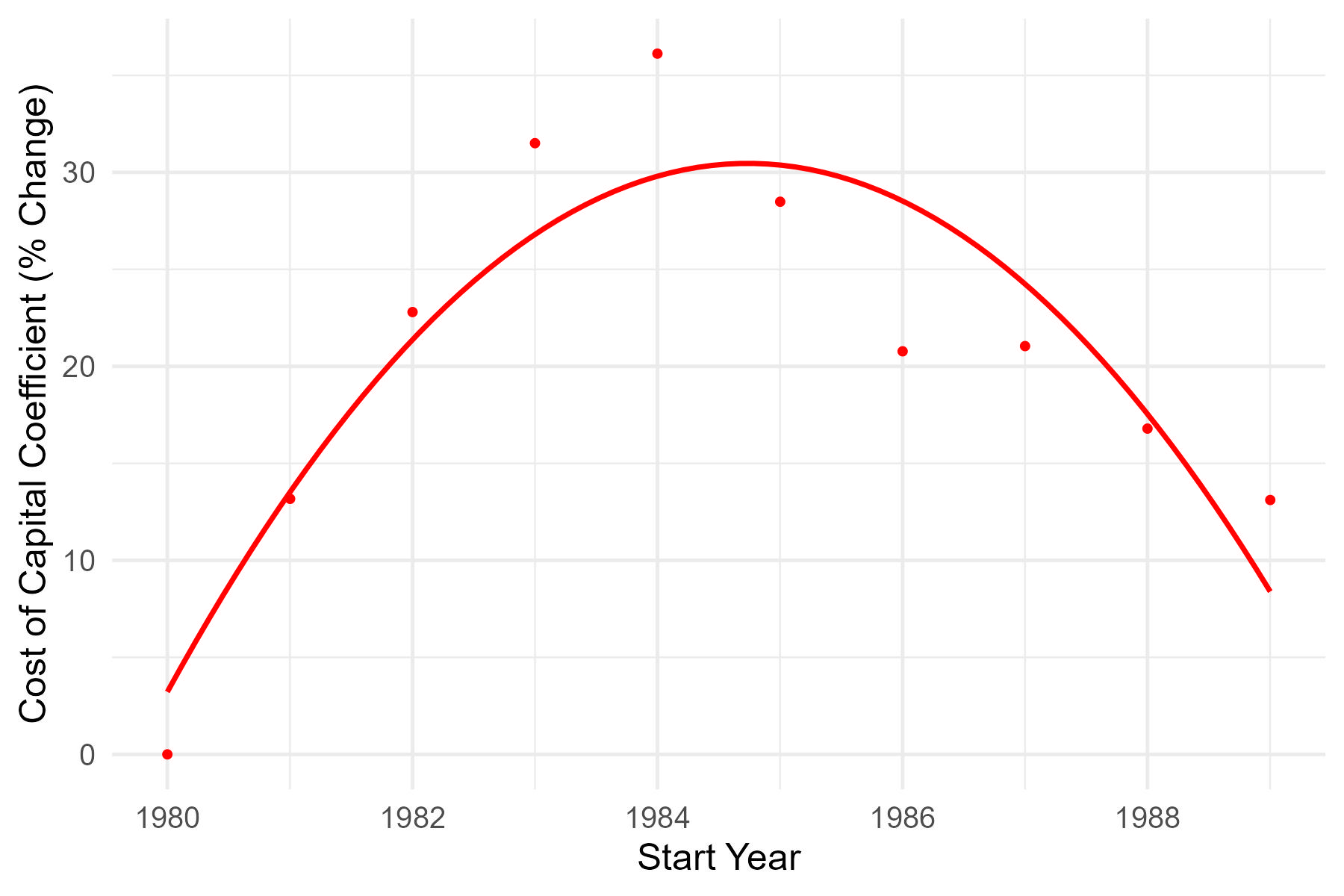}
        \caption{Estimated Trend}
        \label{fig:fig2b}
    \end{subfigure}
    \caption{Cost of Capital: $R_c$. The cost of capital, otherwise known as the weighted average cost of capital (WACC) is composed of the equity and debt costs of capital measured as the weighted average of Moody's Baa bond yield and the 10-year Treasury yield plus a 5\% risk premium. Shown at quarterly frequency.}
    \label{fig:fig2}
\end{figure}

Figure \ref{fig:fig2a} displays the cost of capital. The portion in red is from 1984 to 2014, the period commonly referenced in the literature. Figure \ref{fig:fig2b} plots the percentage change in the linear trend of the cost of capital, starting from different years. We fit a quadratic function over these points.

The trend for the cost of capital is steepest when starting the series in 1984. Data points from the beginning of the sample are highly influential. The cost of capital was 7 pp lower in 1980 than in 1984. Comparing the endpoints of the sample, the cost of capital was nearly the same in 1980 and 2019. 

The economic profit share measures earnings in excess of production costs, including the cost of capital. It's fundamentally a residual after subtracting labor and capital payments from value-added: 
    $$\Pi = 1 - \frac{WL}{Y} - \frac{R_cK}{Y}.$$
Following \citet{basu2019}, we can convert this measure of market power to an implied markup on gross output, as in \citet{de2020rise}. Assuming constant returns to scale and an intermediate input share of revenue of 0.5, we have:
    $$\mathcal{M} = \frac{2}{2 - \Pi}.$$

\begin{figure}[!htbp]
    \centering
    \begin{subfigure}[t]{0.49\textwidth}
        \includegraphics[width=3.25in]{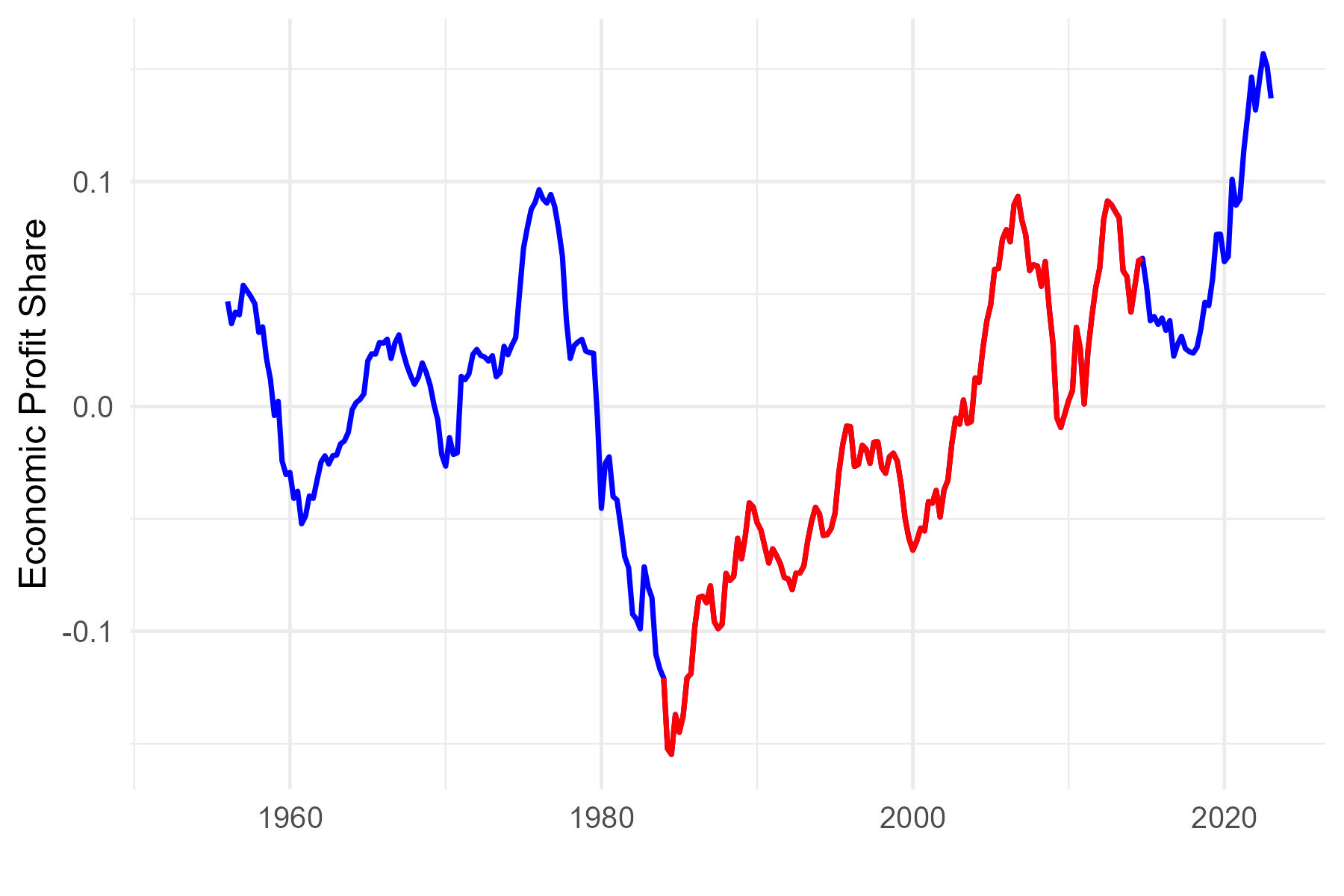}
        \caption{Economic Profit Share}
        \label{fig:fig3a}
    \end{subfigure}
    \begin{subfigure}[t]{0.49\textwidth}
        \includegraphics[width=3.25in]{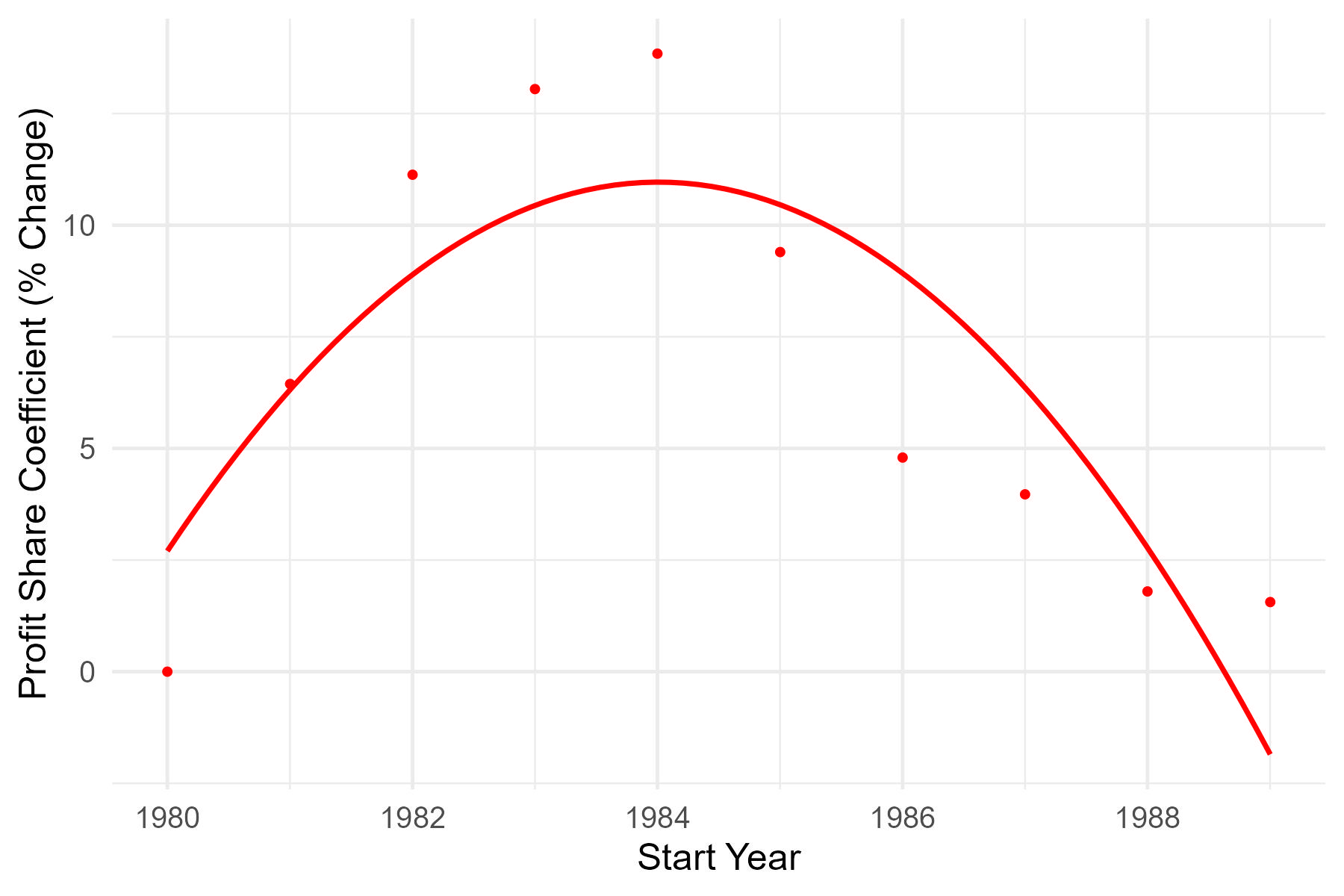}
        \caption{Estimated Trend}
        \label{fig:fig3b}
    \end{subfigure}
    \begin{subfigure}[t]{0.49\textwidth}
        \includegraphics[width=3.25in]{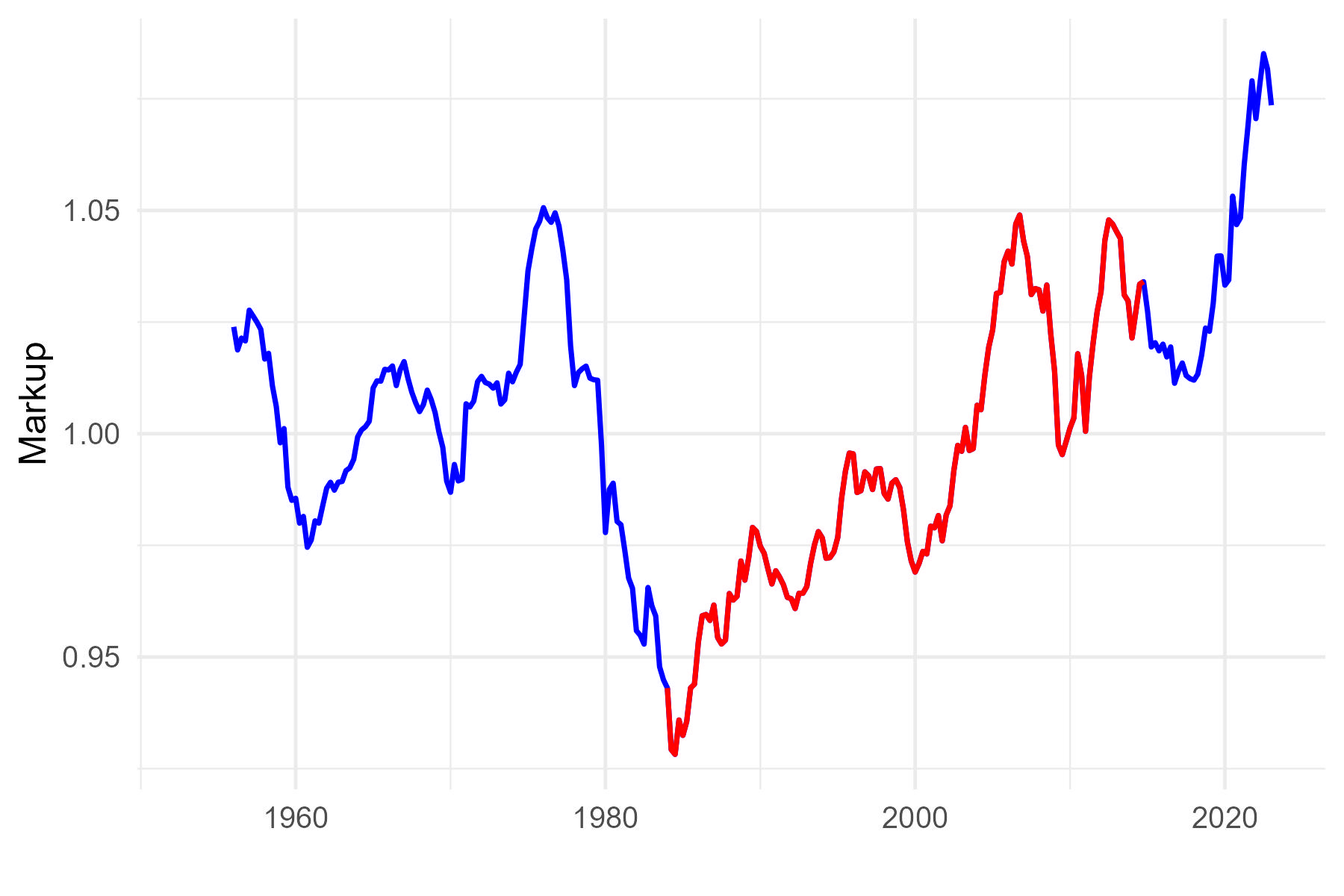}
        \caption{Implied Markup on Gross Output}
        \label{fig:fig3c}
    \end{subfigure}
    \begin{subfigure}[t]{0.49\textwidth}
        \includegraphics[width=3.25in]{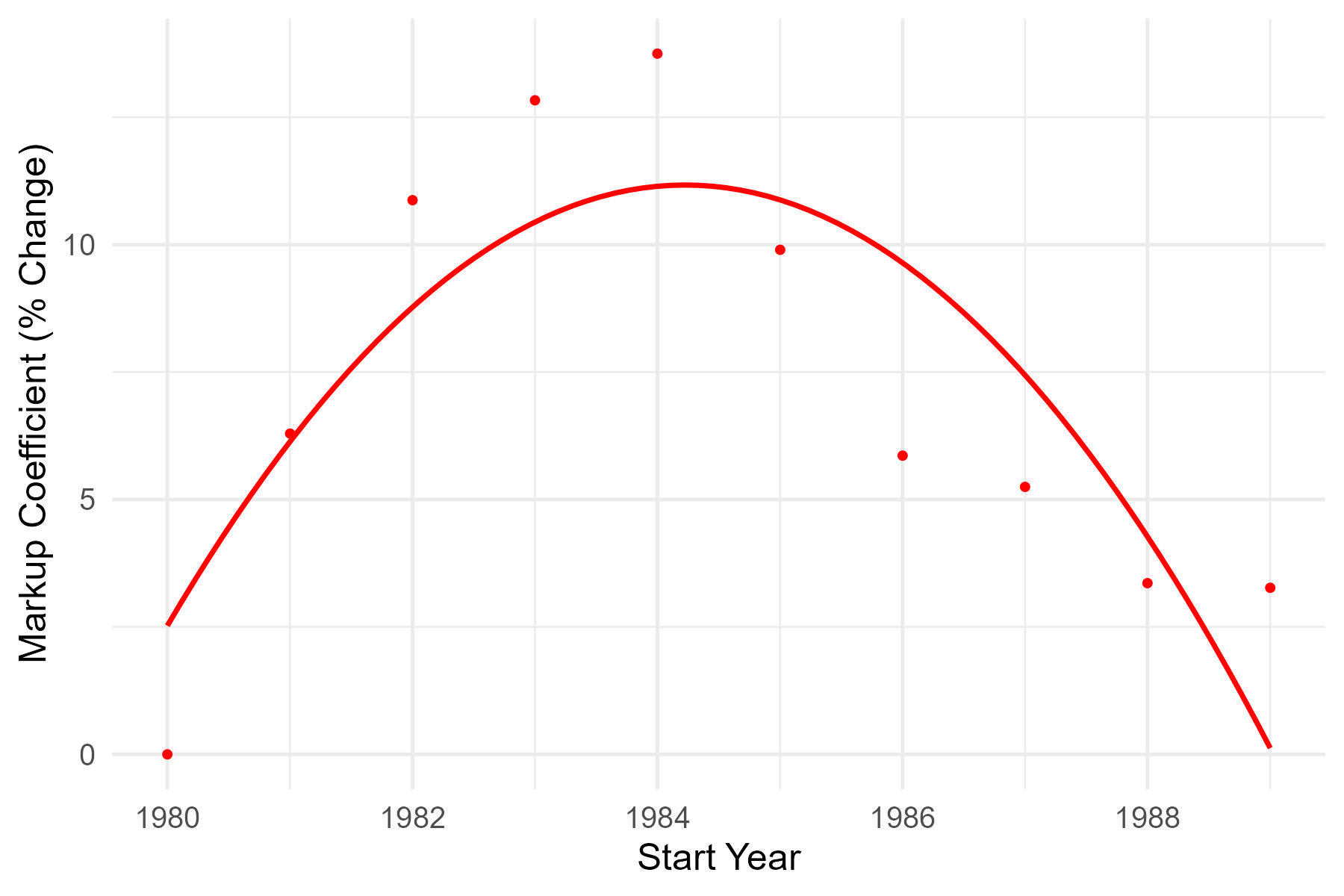}
        \caption{Estimated Trend}
        \label{fig:fig3d}
    \end{subfigure}
    \caption{Economic Profit: $\Pi$, Markup: $\mathcal{M}$. Economic profit is measured as gross value added less compensation of employees less capital costs less taxes on production. Shown at quarterly frequency.}
    \label{fig:fig3}
\end{figure}

Figures \ref{fig:fig3a} and \ref{fig:fig3c} display the economic profit share and implied markups on gross output. Figures \ref{fig:fig3b} and \ref{fig:fig3d} plot the percentage change in the linear trend of the real interest rate and implied markups, starting from different years. Again, we fit a quadratic function over these points. 

The trends for profits and markups are steepest when starting the series in 1984. The data from the beginning of the sample are influential. Economic profits were 4 pp higher in 1980 than in 1984.

\section{Time-Series Volatility and Influence}

The secular trends of rising profits and markups are due to variation in real interest rates. However, real interest rates are notoriously volatile, and their fluctuations can significantly impact our understanding of the economic environment. This issue is particularly salient with financial-market time series, where the data are commonly volatile. In this section, we conclude with a discussion of statistical influence. 

Influence provides an analytic formula that shows how an estimator changes when a single data point is included or excluded, and has been applied in many areas of econometrics \citep{erickson2000measurement, andrews2017measuring, ichimura2022influence}.

    $$\frac{t - \mathbb{E}[t]}{\mathrm{Var}[t]}(y - \mathbb{E}[y] - \beta(t - \mathbb{E}[t] ))$$

The formula indicates points far from the average time value have more impact on the regression, which is mechanically highest for points at the beginning or end of the sample. Moreover, points that are far from the average value of the series, as often with high volatility in a time series, will also have more influence. The influence of a data point is even more pronounced when the estimated beta is far from zero, which is exactly the case in studies that argue for rises or falls in secular trends. Therefore, the choice of the sample start (and end dates) is itself influential; it can significantly alter the conclusions drawn from the regression.

\begin{figure}[!ht]
    \centering
    \begin{subfigure}[t]{0.49\textwidth}
        \includegraphics[width=3.25in]{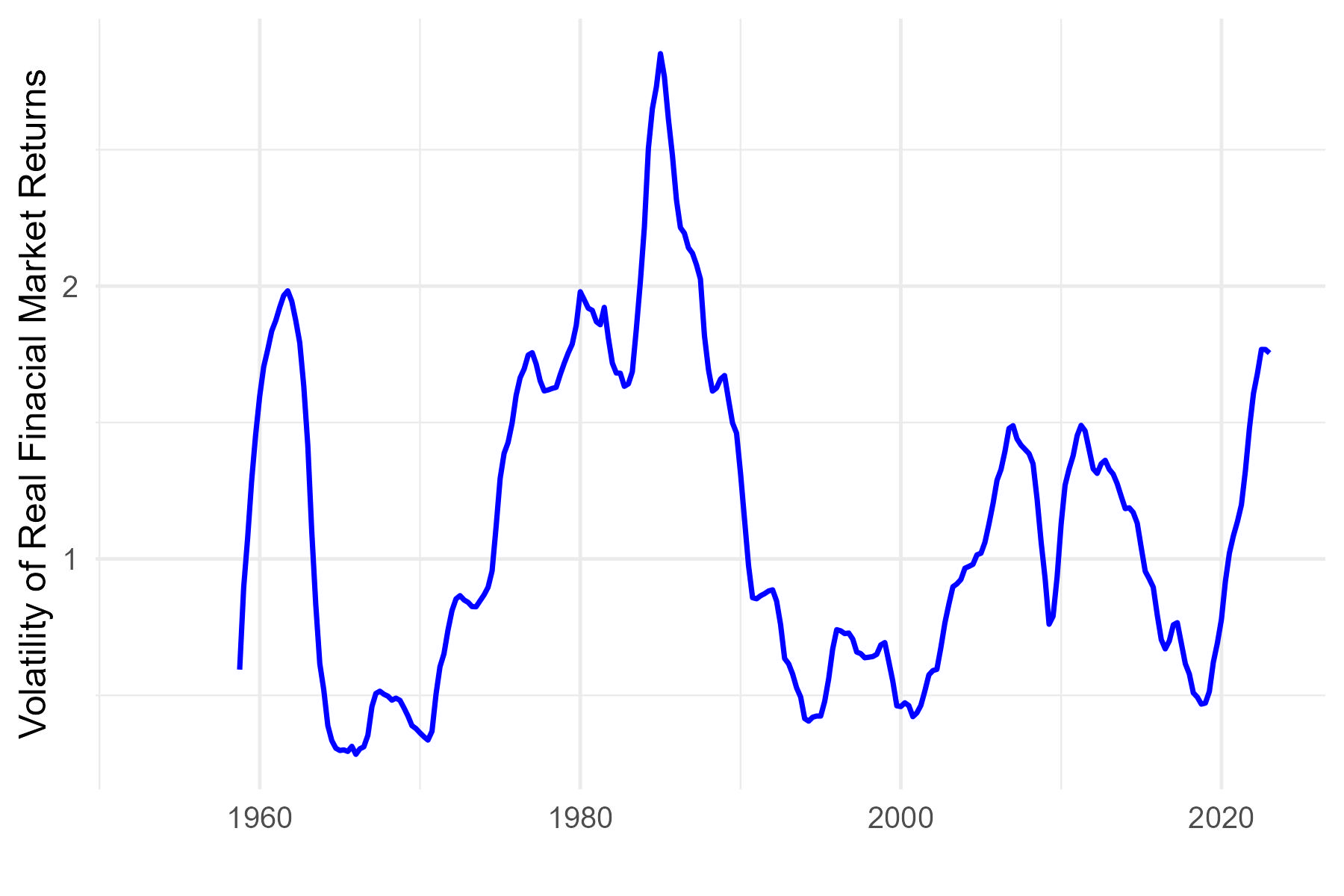}
        \caption{Volatility of Interest Rate}
        \label{fig:fig4a}
    \end{subfigure}
    \begin{subfigure}[t]{0.49\textwidth}
        \includegraphics[width=3.25in]{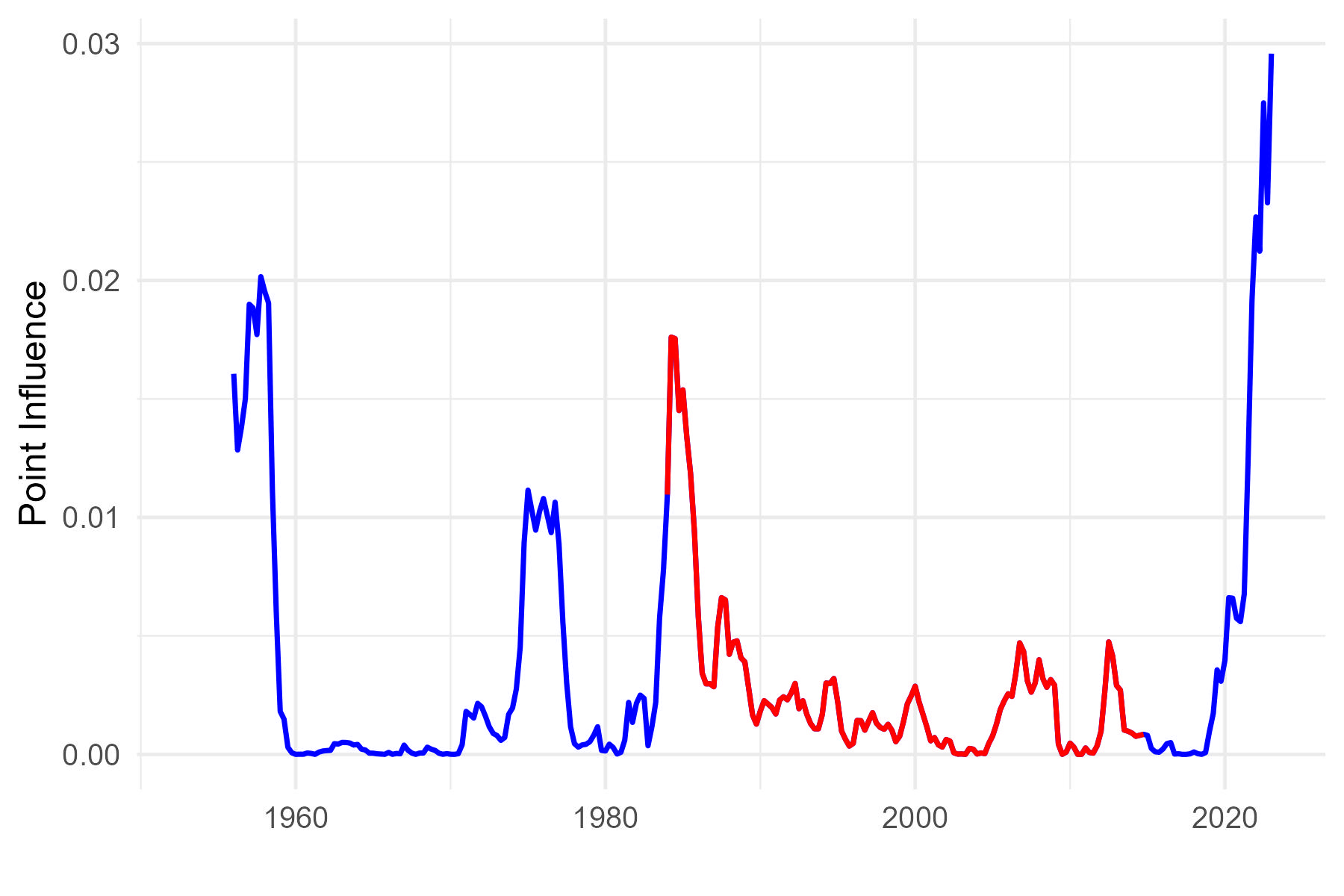}
        \caption{Cook's Distance}
        \label{fig:fig4b}
    \end{subfigure}
    \caption{Statistical Influence. Volatility is measured as the rolling standard deviation of the real interest rate from figure \ref{fig:fig1} over a window of 5 year window. Cook's distance is calculated the footnote 4. Shown at quarterly frequency.}
\end{figure}

Figure \ref{fig:fig4a} plots the volatility of the real interest rate, measured as the standard deviation in a 5-year rolling window. Figure \ref{fig:fig4b} plots the Cook's distance of each data point. Cook's distance is a common metric of statistical influence in univariate data and summarizes the change in regression coefficients when observation $i$ is removed from the sample.\footnote{Explicitly for simple regression $D_i = \frac{\sum_{j=1}^{n}(\hat y_j - \hat y_{j(i)})^2}{2(MSE)}$ where $\hat y_j$ is the $j$ fitted response value, $\hat y_{j(i)}$ is the $j$ fitted response value excluding observation $i$ and $MSE$ is the mean-squared error.} The secular trends in profits and markups (Figure \ref{fig:fig3}) and the cost of capital (Figure \ref{fig:fig2}) are most exaggerated when plotting across endpoints with the highest real interest rate volatility and influence. Moving across these high-volatility years dramatically changes the trend slope. Coinciding with a spike in volatility after 2019, the economic profit share is at record levels of 15 pp in 2021 and 2022.

To illustrate the effect of time-series volatility on sample point-to-point estimates, Table \ref{table:tbl1} collects the 15-year-long differences of our main variables by sample stop year. As volatility declines in the sample, the long difference in interest rates, the cost of capital, profits, and markups nearly vanish. However, after 2019, a spike in the volatility of real the interest rate increases the difference by a factor of 3. These endpoints also disproportionately influence the time series trend seen in the spike of point influence after 2020 in Figure \ref{fig:fig4b}.

\begin{table}[!htbp] \centering 
    \input{Tables/table.tex}
    \caption{15 year long differences by sample end date. Table shows the differences between point estimates for the real interest rate, the cost of capital, profit rate, and markups over a shifting 15 year window. } 
    \label{table:tbl1} 
\end{table}

Volatility in financial-market rates poses a challenge to accurately estimating economic profits and markups. The prevailing methodology for calculating economic profits and implied markups takes into account payments to labor and capital. However, these methodologies often overlook the volatility inherent in financial-market rates compared to more stable returns on capital. Such volatility in capital returns can significantly skew calculated markups and profits, thereby affecting observed trends. Consequently, the notable shifts in profits and markups around 1984 and 2022 are influenced by fluctuations in the cost of capital, which factors like changes in expected capital inflation may drive.

\section{Concluding Implications}

While existing research commonly attributes rising economic profits and markups to broader secular trends, our study adds an important qualifier. We demonstrate the choice of sample start date can significantly influence the size of these trends, particularly when the time series is volatile, as is often the case with financial-market data. We remark our paper focuses only on the role of sample start dates, but analogous arguments apply to sample end dates. In our worked example, extending the sample backward by just 4 years explains 25\% of the increase in economic profits over since the 1980s, or \$3000 per worker in 2019.

Our analysis implies even small differences in sample selection can lead to large differences in trend estimation. Given communicating research inevitably loses some detail, it's key that takeaways are not contingent on something as simple as a sample start date. Therefore, we conclude, researchers should justify their choice of time-series sample selection. We also advise researchers to incorporate sensitivity checks to keep these important statistics reliable and portable for future use.

\bibliography{bibliography}

\end{document}

%% file: Tables/table.tex

\begin{tabular}{@{\extracolsep{7pt}} cccccc} 
\\[-1.8ex]\hline 
\hline \\[-1.8ex] 
Sample & $\Delta (i- \nu)$ & $\Delta  R_c$ & $\Delta  \Pi$  & $\Delta \mathcal{M}$ \\ 
\hline \\[-1.8ex] 
$1997-2012$ & $-3.65$ & $-3.62$ & $7.58$        & 0.39 \\ 
$1999-2014$ & $-3.95$ & $-4.28$ & $10.35$ & 0.52 \\ 
$2001-2016$ & $-2.54$ & $-2.45$ & $8.40$ & 0.042 \\
$2003-2018$ & $-0.41$ & $-0.61$ & $2.90$ & 0.014 \\
$2005-2020$ & $-1.78$ & $-1.64$ & $ 2.06$ & 0.011 \\ 
$2007-2022$ & $-2.32$ & $-2.33$ & $6.64$ & 0.037 \\ 
\hline \\[-1.8ex] 
\end{tabular}

%% file: AEL.bbl
\begin{thebibliography}{}

\bibitem[\protect\citeauthoryear{Andrews, Gentzkow, and Shapiro}{Andrews
  et~al.}{2017}]{andrews2017measuring}
Andrews, I., M.~Gentzkow, and J.~Shapiro (2017).
\newblock Measuring the sensitivity of parameter estimates to estimation
  moments.
\newblock {\em Quarterly Journal of Economics\/}.

\bibitem[\protect\citeauthoryear{Autor, Dorn, Katz, Patterson, and
  Van~Reenen}{Autor et~al.}{2020}]{autor2020fall}
Autor, D., D.~Dorn, L.~Katz, C.~Patterson, and J.~Van~Reenen (2020).
\newblock The fall of the labor share and the rise of superstar firms.
\newblock {\em Quarterly Journal of Economics\/}.

\bibitem[\protect\citeauthoryear{Barkai}{Barkai}{2020}]{barkai2020declining}
Barkai, S. (2020).
\newblock Declining labor and capital shares.
\newblock {\em Journal of Finance\/}~{\em 75\/}(5).

\bibitem[\protect\citeauthoryear{Basu}{Basu}{2019}]{basu2019}
Basu, S. (2019).
\newblock Are price-cost markups rising in the {U}nited {S}tates? {A}
  discussion of the evidence.
\newblock {\em Journal of Economic Perspectives\/}.

\bibitem[\protect\citeauthoryear{Blanchard}{Blanchard}{2019}]{blanchard2019public}
Blanchard, O. (2019).
\newblock Public debt and low interest rates.
\newblock {\em American Economic Review\/}~{\em 109\/}(4).

\bibitem[\protect\citeauthoryear{Card and Krueger}{Card and
  Krueger}{1995}]{card1995time}
Card, D. and A.~Krueger (1995).
\newblock Time-series minimum-wage studies: A meta-analysis.
\newblock {\em American Economic Review\/}.

\bibitem[\protect\citeauthoryear{{Council of Economic Advisors}}{{Council of
  Economic Advisors}}{2023}]{cea2023}
{Council of Economic Advisors} (2023).
\newblock Protecting competition through updated merger guidelines.
\newblock Issue brief, White House.

\bibitem[\protect\citeauthoryear{Covarrubias, Guti{\'e}rrez, and
  Philippon}{Covarrubias et~al.}{2020}]{covarrubias2020good}
Covarrubias, M., G.~Guti{\'e}rrez, and T.~Philippon (2020).
\newblock From good to bad concentration? {US} industries over the past 30
  years.
\newblock {\em NBER Macroeconomics Annual\/}.

\bibitem[\protect\citeauthoryear{Davis, Sollaci, and Traina}{Davis
  et~al.}{2023}]{davis2023profit}
Davis, C., A.~Sollaci, and J.~Traina (2023).
\newblock Profit puzzles and the fall of public-firm profit rates.
\newblock {\em Kelley School of Business Research Paper\/}.

\bibitem[\protect\citeauthoryear{De~Loecker, Eeckhout, and Unger}{De~Loecker
  et~al.}{2020}]{de2020rise}
De~Loecker, J., J.~Eeckhout, and G.~Unger (2020).
\newblock The rise of market power and the macroeconomic implications.
\newblock {\em Quarterly Journal of Economics\/}.

\bibitem[\protect\citeauthoryear{Edmond, Midrigan, and Xu}{Edmond
  et~al.}{2023}]{edmond2023costly}
Edmond, C., V.~Midrigan, and D.~Y. Xu (2023).
\newblock How costly are markups?
\newblock {\em Journal of Political Economy\/}~{\em 131\/}(7).

\bibitem[\protect\citeauthoryear{Eggertsson, Robbins, and Wold}{Eggertsson
  et~al.}{2021}]{eggertsson2021kaldor}
Eggertsson, G., J.~Robbins, and E.~G. Wold (2021).
\newblock Kaldor and {P}iketty’s facts: The rise of monopoly power in the
  {U}nited {S}tates.
\newblock {\em Journal of Monetary Economics\/}.

\bibitem[\protect\citeauthoryear{Erickson and Whited}{Erickson and
  Whited}{2000}]{erickson2000measurement}
Erickson, T. and T.~Whited (2000).
\newblock Measurement error and the relationship between investment and q.
\newblock {\em Journal of Political Economy\/}~{\em 108\/}(5).

\bibitem[\protect\citeauthoryear{Farhi and Gourio}{Farhi and
  Gourio}{2018}]{farhi2018accounting}
Farhi, E. and F.~Gourio (2018).
\newblock Accounting for macro-finance trends: Market power, intangibles, and
  risk premia.
\newblock {\em Brookings Papers on Economic Activity\/}~(2).

\bibitem[\protect\citeauthoryear{Hall}{Hall}{2018}]{hall2018new}
Hall, R. (2018).
\newblock New evidence on the markup of prices over marginal costs and the role
  of mega-firms in the {US} economy.
\newblock Working Paper 24574, National Bureau of Economic Research.

\bibitem[\protect\citeauthoryear{Hall and Jorgenson}{Hall and
  Jorgenson}{1967}]{hall1967tax}
Hall, R. and D.~Jorgenson (1967).
\newblock Tax policy and investment behavior.
\newblock {\em American Economic Review\/}, 391--414.

\bibitem[\protect\citeauthoryear{Ichimura and Newey}{Ichimura and
  Newey}{2022}]{ichimura2022influence}
Ichimura, H. and W.~Newey (2022).
\newblock The influence function of semiparametric estimators.
\newblock {\em Quantitative Economics\/}~{\em 13\/}(1).

\bibitem[\protect\citeauthoryear{Jensen, Kelly, and Pedersen}{Jensen
  et~al.}{2022}]{jensen2022there}
Jensen, T., B.~Kelly, and L.~Pedersen (2022).
\newblock Is there a replication crisis in finance?
\newblock {\em Journal of Finance\/}.

\bibitem[\protect\citeauthoryear{Karabarbounis and Neiman}{Karabarbounis and
  Neiman}{2019}]{karabarbounis2019accounting}
Karabarbounis, L. and B.~Neiman (2019).
\newblock Accounting for factorless income.
\newblock {\em NBER Macroeconomics Annual\/}~{\em 33\/}(1).

\bibitem[\protect\citeauthoryear{Philippon}{Philippon}{2019}]{philippon2019causes-key}
Philippon, T. (2019).
\newblock Causes, consequences, and policy responses to market concentration.
\newblock Technical report, Aspen Economic Strategy Group.

\bibitem[\protect\citeauthoryear{Shapiro}{Shapiro}{2018}]{shapiro2018antitrust}
Shapiro, C. (2018).
\newblock Antitrust in a time of populism.
\newblock {\em International Journal of Industrial Organization\/}~{\em 61}.

\bibitem[\protect\citeauthoryear{Traina}{Traina}{2018}]{traina2018aggregate}
Traina, J. (2018).
\newblock Is aggregate market power increasing? {P}roduction trends using
  financial statements.
\newblock Working Paper~17, Stigler Center.

\bibitem[\protect\citeauthoryear{White}{White}{2000}]{white2000reality}
White, H. (2000).
\newblock A reality check for data snooping.
\newblock {\em Econometrica\/}~{\em 68\/}(5).

\end{thebibliography}
